\documentstyle[twoside,fleqn,npb,epsfig]{article}
\newcommand{\AmS}{{\protect\the\textfont2
  A\kern-.1667em\lower.5ex\hbox{M}\kern-.125emS}}

\title{Future Polarised DIS Fixed Target Experiments}

\author{E.M. Kabu{\ss}\address{Institut f\"ur Kernphysik, 
Universit\"at Mainz,\\
Becherweg 45, D 55099 Mainz}%
        \thanks{Supported by the BMBF}}

\begin{document}

\begin{abstract}
New experiments in polarised deep inelastic scattering will mainly concentrate
on the measurement of semi-inclusive asymmetries. Especially,
the upgraded HERMES experiment at DESY and the newly build COMPASS experiment
at CERN will investigate the gluon polarisation via open charm and high
$p_{\rm T}$ hadron pair production, study in detail the flavour decomposition
of the quark helicity distributions and measure the tranversity distributions
 with tranversely polarised targets.
\end{abstract}

\maketitle

\section{INTRODUCTION}

The spin structure of the nucleon has been investigated in polarised deep 
inelastic scattering by a series of experiments at CERN, SLAC and 
DESY \cite{emk_roland}. These
experiments were initiated by the discovery of the EMC in 1987 that the 
contribution of the quark helicities to the proton spin is much smaller than 
expected originally \cite{emk_emc}. The new experiments confirmed the original finding of the
EMC also for the neutron that
the singlet axial vector current matrix element is about 1/2 to 1/3 of the
predicted value of 0.6 \cite{emk_ej}.

Up to now only the contribution of quark helicities to the nucleon spin
was studied.
Further insight into the spin structure of the nucleon can be gained by
investigating the gluonic contribution and the contribution of angular
momentum. In addition a measurement of the decomposition of the quark 
contribution into the different quark flavours will yield a deeper 
understanding of the nucleon spin puzzle.

There is a whole list of topics which need more detailed studies:
\begin{itemize}
\item The flavour decomposition of the polarised quark distributions
can be extracted via the measurement of semi-inclusive asymmetries. 
\item The gluon polarisation can be measured with the help of open charm 
production or high $p_{\rm T}$ hadron pairs.
\item Polarised fragmentation functions and spin transfer can be studied by 
measuring the $\Lambda$ polarisation in the current and target fragmentation
region.
\item A new topis is the study of transversity in scattering on a transversely
polarised target using the Collins effect. First signal were presented by
SMC and HERMES during this workshop \cite{emk_trans1,emk_trans2}.
\item The total angular momentum of quarks might be investigated via deeply
virtual Compton scattering.
\end{itemize}
Further topics on the list refer to the measurement of off-forward parton
distributions, vector meson production etc.

The common feature of all these new measurements is the need to detect one or 
several hadrons in addition to the scattered lepton.

\section{MEASUREMENT OF $\Delta G$}

Up to now the gluon polarisation has been investigated by indirect methods 
using NLO analyses of
structure function data \cite{emk_roland}. They indicate that integral $\Delta G$ is positive 
and of the order of 1 at $Q^2=1$~GeV$^2$ with fairly large errors while
the functional form of $\Delta G(x)$ is completely unknown although there
are some prejudices that $\Delta G(x)$ is largest around $x\approx 0.1$.

\begin{figure}[htb]
\begin{center}
\epsfig{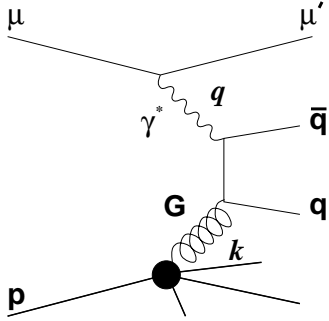}
\caption{\label{emk_openc} The photon fusion diagramm.}
\end{center}
\end{figure}

The cleanest channel for a direct measurement of $\Delta G$ in polarised
DIS is the photon gluon fusion process (PGF) as it depends in leading order
on the gluon distribution (see fig.~\ref{emk_openc}). In this context several
methods are being discussed to extract $\Delta G/G$:
\begin{itemize}
\item Open charm procduction: $$\gamma {\rm N} \rightarrow {\rm c \bar c X}
\rightarrow {\rm D^0 X}$$
\end{itemize}
PGF is signalled by the dectection of charmed particles in the final state,
especially by ${\rm D^0}$ and $\Lambda_{\rm c}$ (close to threshold).
The ${\rm D^0}$'s are reconstructed via their decay to e.g. ${\rm K}\pi$
or $\mu {\rm K}\pi$. This process should yield a clear signal directly related
to $\Delta G$. Here, the hard scale is given by $2{\rm m_c}$.

The cross section for open charm production $\sigma ^{\gamma {\rm N}
\rightarrow {\rm c\bar c}}$ is large for quasi real photons and 
$\Delta \sigma ^{\gamma {\rm N}
\rightarrow {\rm c\bar c}}/\sigma ^{\gamma {\rm N}
\rightarrow {\rm c\bar c}}$ is largest for photon energies between 30 and 
80~GeV. The asymmtry $a_{\rm LL}$ for the hard subprocess $\gamma {\rm
  g}\rightarrow {\rm c\bar c}$ is about 1 at threshold ($2{\rm m_c}$)
while $a_{\rm LL}=-1$ for large energies. In this case a positive gluon
polarisation will leaad to a negative photon nucleon asymmetry:
$$A_{\gamma {\rm N}}^{\rm c\bar c}= \langle a_{\rm LL} \rangle \langle
{\Delta G}/{G}\rangle\,.$$
\begin{itemize}
\item Hidden charm production:
 $$\gamma {\rm N} \rightarrow {\rm c \bar c X}
\rightarrow {\rm J}/\psi{\rm X}$$
\end{itemize}
Here the photon gluon process is signalled by the production of a ${\rm
  J}/\psi$ which is identified by its decay into a muon pair. While this is a
very clean experimental signal the cross section is reduced considerably
compared to open charm production. Moreover the relation of the signal to 
 ${\Delta G}/{G}$ has to be done via the colour singlet or the colour octet
model, a question which is not yet settled in unpolarised DIS.
\begin{itemize}
\item High $p_{\rm T}$ hadron pairs:  
$$\gamma {\rm N} \rightarrow {\rm q \bar q    X} \rightarrow 2{\rm jets~X}$$
\end{itemize}
This third method tries to select all PGF events not only the ${\rm c\bar  c}$
production. The transverse momenta of the produced jets give the necessary
hard scale.

At the moderate energies of fixed target experiments jets are not available
but one can use fast hadrons instead \cite{emk_bravar}. Selecting 
oppositely charged high 
$p_{\rm T}$ hadron pairs  will enhance PGF events, but there is a considerable
background especially from the QCD Compton process. Thus the measured 
asymmetry is given by
$$A_{\mathrm LL}^{\mathrm HH}\approx  \langle a_{\rm LL}^{\rm PGF} \rangle
\langle \frac{\Delta G}{G}\rangle \frac{\sigma^{\mathrm PGF}}
{\sigma^{\mathrm tot}} + \langle a_{\rm LL}^{\rm COM} \rangle \langle 
\frac{\Delta u}{u}\rangle \frac{\sigma^{\mathrm COM}}{\sigma^{\mathrm tot}}\,.
$$
The hard asymmtries $a_{\rm LL}^{\rm PGF}$ and $a_{\rm LL}^{\rm COM}$ are
large in the $Q^2$ range of the fixed target experiments and have opposite
signs. 

In addition one has to investigate the contribution due to resolved
photons. Thus, the results from this method will be dominated by 
systematic effects due to large background subtractions. A first attempt to use
the method was presented by the HERMES collaboration during this workshop
\cite{emk_hermesglue}.

\section{FACILITIES}
Up to now the experiments concentrated on inclusive measurements of the spin
structure functions $g_1$ and $g_2$. This era comes to an end with the present
E155X experiment at SLAC \cite{emk_e155x} where data are being taken for a
precise measurement of $g_2^{\rm p,d}$. This effort will be continued at 
Jefferson Lab \cite{emk_cebaf}, where a high statistics
measurement of the large $x$ behaviour of $g_1$ and $g_2$ is being planned
using a polarised $^3{\rm He}$ target.

Several facilities will be available during the next years to measure 
semi-inclusive properties in polarised DIS:
\begin{itemize}
\item HERMES at DESY, which is in full swing measuring semi-inclusive
  asymmetries, has started a large upgrade program to attack several of the
questions listed above.
\item The COMPASS experiment is being setup at the CERN 100--200~GeV muon
beam and will start data taking in 2000 focussing in the beginning on a
measurement of $\Delta G$.
\item At MAMI in Mainz, ELSA in Bonn and Jefferson Lab measurements of the 
GDH sumrule and the generalized GDH sumrule will be continued.
\end{itemize}
In addition there are plans for future high luminosity maschines where a
continuation and extension of the present spin program will be feasible,
e.g. the ELFE proposal at CERN or DESY to study polarised DIS and the 
APPOLON at ELFE and the SLAC real photon beam proposal to investigate 
photoproduction.

In the following I will concentrate on the HERMES upgrade and the COMPASS
experiment.

\section{HERMES UPGRADE PROGRAM}

The main aims of the upgrade program are
\begin{itemize}
\item Particle identification in the full hadron momentum range, i.e.
pion, kaon and proton separation,
\item Enlarged muon acceptance and improved muon identification,
\item Electron acceptance at very small scattering angles,
\item Enlarged hadron acceptance covering also negative $x_{\rm F}$.
\end{itemize}
The first item was attacked with the installation of a dual RICH \cite{emk_rich}. 
Due to the
combination of an aerogel with a gas radiator $\pi/{\rm K/p}$ separation
is achieved in the full momentum range up to 20~GeV. To yield high precision
measurements of the Cerenkov rings the RICH is being red by an array of
photomultiplier. 

The RICH is already installed and was succesfully operated in 1998. Currently
particle identification is being implemented in the analysis chain.

The new muon filter system has been installed during the last shutdown
\cite{emk_charmupr}. It
consists out of an iron absorber at the end of the spectrometer followed by a
scintillator hodoscope.

The enlarged muon acceptance (for scattering angles above 170~mrad) will be 
made available by using tracks passing part of the magnet yoke. During the 
shutdown in May 1999 additional scintillators will be installed covering the region
between 140 and 270~mrad behind the spectrometer magnet.

With a forward quadrupole spectrometer \cite{emk_charmupr} the electron 
acceptance
will be extended to smaller scattering angles. This is especially important
for quasireal photoproduction events. Up to now only 10\% of the scattered 
electrons were detected by the luminosity monitor. The new spectrometer
will add another 16\%.

For this, quadrupoles with larger apertures were installed in the last long
shutdown. The electrons will be measured using a small vertical driftchamber
installed inside the quadrupole.
After the successfull test of a prototype chamber the system will be installed
in May 1999. 

The next topic is the enlarged hadron acceptance \cite{emk_lambda}. 
For this purpose a wheel of 
silicon detectors is being constructed to be positioned right after the target
cell. This will enlarge the acceptance to $x_{\rm F} < 0$. Monte Carlo
simulations showed that this improvement is especially important for measuring
the $\Lambda$ decay products. The installation of the system will start in May
1999.

The last project on the list is the recoil detector. It will consist out of a
layer of double sided silicon detectors positioned below the target cell. It
will be used to measure recoil particles from the target to identify
diffractive events and measure tagged structure functions. This year a
prototype detector was tested successfully. The installation of the full
system is forseen for 2001 provided funding is available.

This upgrade will allow
\begin{itemize}
\item To study the flavour decomposition in more detail, e.g. measure strange 
quark polarisation $\Delta s(x)/s(x)$.
\item A measurement of the gluon polarisation with several methods.
Using open charm production and the ${\rm D_0}$ decay into $\pi{\rm K}$ and
$\mu\pi{\rm K}$ a precision of $\Delta G/G$ of 0.44 rsp. 0.40 can be reached
with a luminosity of 80~pb$^{-1}$. The measurement of hidden charm yields
$\delta(\Delta G/G)$ of about 0.69.
In addition the enlarged hadron acceptance will improve the measurement via
high $p_{\rm T}$ hadron pairs.
\item The measurement of the $\Lambda$ polarisation in the current fragmentation
region will yield a significant measurement of the polarised fragmentation
function $\Delta D^{\Lambda}_{\rm u}/D^{\Lambda}_{\rm u}$, while the spin
correlation for strange quarks will be studied in the target fragmentation
region. 
\item After 2000, measurements with a transversely polarised target will allow
studies of azimuthal spin asymmetries to extract tranversity distributions.
\end{itemize}

\section{THE COMPASS EXPERIMENT}

Currently, the COMPASS experiment is being setup at the CERN M2 muon beam line
to study polarised deep inelastic muon nucleon scattering. In addition a 
hadron program is 
planned e.g. to study charmed baryons and search for glue balls 
\cite{emk_compass}.

\begin{figure*}[htb]
\newlength{\digitwidth} \settowidth{\digitwidth}{\rm 0}
\catcode`?=\active \def?{\kern\digitwidth}
\begin{center}
\epsfig{file=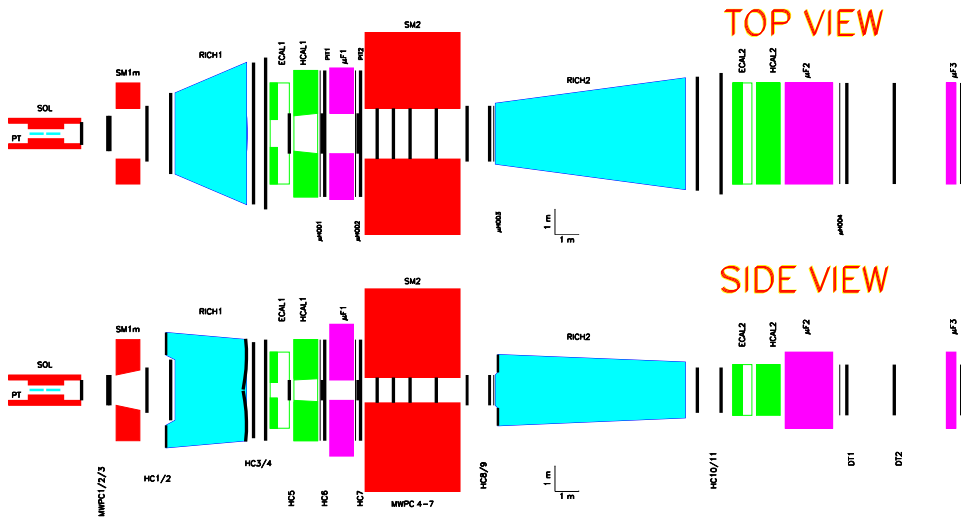,angle=270,width=10cm}
\caption{\label{emk_spectr} Schematic top and side view of the 
COMPASS spectrometer}
\end{center}
\end{figure*}

Compared to the previous muon experiment from SMC COMPASS will have an
increased muon beam intensity of $2\cdot 10^8\mu$/14.4~s with 100--200 GeV
and 80\% polarisation. Together with two oppositely polarised target cells of
60~cm length filled with $^6$LiD or NH$_3$ a luminosity of about 2~fb$^{-1}$
per year can be reached.

The spectrometer is optimized for large hadron acceptance and particle
identification (see fig.~\ref{emk_spectr}). To achieve acceptance up to $\pm180$~mrad a new target
solenoid is being constructed with a large opening minimizing the multiple 
scattering for hadron tracks at large angles.

The spectrometer itself consists out of two stages. Each stage has a dipole
spectrometer magnet surrounded by tracking chambers followed by a RICH
detector,
an electromagnetic and a hadronic calorimeter and a muon filter system.

A special feature of the muon beam
is the 10\% halo of muon tracks surrounding the muon beam up to a radius of
0.5 to 1~m. In addition, scattered muons have to be
detected very close and in the muon beam for the measurement of quasireal
photoproduction. Thus the tracking system has to be split into three regions. 

The beam and scattered muons with very small angles will be measured by
scintillating fiber hodoscopes.
Small angle tracking will be performed with micromega chambers (stage 1)
and GEM detectors (stage 2) while drift (stage 1), proportional
(stage 2) and straw chambers will be used for the large angle tracking.

For the first year of data taking the detector will not be complete,
especially
the RICH in stage 2, the electromagnetic calorimeter electronics and part of
the large angle tracking will be missing. Thus the measurements will
concentrate on $\Delta G/G$ via quasireal photoproduction.

Using the above mentioned luminosity 82k charm events are expected with 1.2 ${\rm D^0}$
per charm event. The ${\rm D^0}$'s will be reconstructed via their $\pi{\rm
  K}$ decays. Due to MCS in the target it is not possible to reconstruct the
decay vertex, thus the large combinatorial background has to be reduced by
strict cuts on the ${\rm D^0}$ kinematics. This should result in 900 charm
events/day with a S:B of 1:3.9. Within 1.5~y with a $^6$LiD target a precision
of $\delta A^{\rm c\bar c}_{\gamma{\rm N}}\approx 0.05$ could be reached
translating to $\delta(\Delta G/G)\approx 0.14$ at $\langle x_{\rm g}
\rangle=0.14$. 
Using e.g. additional decay channels or ${\rm D^{\star}}$ tagging will improve 
the  results. 

Alternatively high $p_{\rm T}$ hadron pairs will be used to extract the gluon
polarisation \cite{emk_bravar}. To reduce the background due to the QCD 
Compton process a series
of cuts (opposite charge, high  $p_{\rm T}>1$~GeV, opposite azimuth,  
$p_{\rm  T}$ balance, positive $x_{\rm F}$) have to be applied resulting 
in a good
signal to background ratio of about 1:1. Due to the suppression of strangeness
in the fragmentation process K$^+$K$^-$ pairs yield an even cleaner signal.

The gluon polarisation can be studied in the range $0.04<x_{\rm g}<0.2$ for 200~GeV
muon energy. With 1~y of data taking a precision of $\delta(\Delta G/G)\approx
0.05$ should be achievable. The error of the gluon polarisation will then be
dominated by systematic effects for this analysis.

With the described spectrometer, especially with the full setup, all topics 
discussed in the introduction can be investigated like
the flavour decomposition of the quark helicity distributions and polarised 
fragmentation
functions. With a tranversely polarised target azimuthal asymmetries will
be measured to extract transversity distributions. The possibility to study 
deeply virtual Compton scattering is currently being investigated.
 
\section{SUMMARY}
 During the next years a rich experimental program is going on in fixed
 target polarised DIS. Experiments at DESY and CERN will do detailed studies
of semi-inclusive processes to unravel more of the details of the nucleon
spin structure.

\end{document}